\documentstyle[amsfonts,twocolumn,aps,graphicx]{revtex}
\draft

\begin{document}
\title{Constructing Infinite Particle Spectra}
\author{O.A.~Castro-Alvaredo and A.~Fring}
\address{Institut f\"{u}r Theoretische Physik, Freie Universit\"{a}t
Berlin, Arnimallee 14, D-14195 Berlin, Germany }
\date{\today}
\maketitle

\begin{abstract}
We propose a general construction principle which allows to include an
infinite number of resonance states into a scattering matrix of hyperbolic
type. As a concrete realization of this mechanism we provide new S-matrices
generalizing a class of hyperbolic ones, which are related to a pair of
simple Lie algebras, to the elliptic case. For specific choices of the
algebras we propose elliptic generalizations of affine Toda field theories
and the homogeneous sine-Gordon models. For the generalization of the
sinh-Gordon model we compute explicitly renormalization group scaling
functions by means of the c-theorem and the thermodynamic Bethe ansatz. In
particular we identify the Virasoro central charges of the corresponding
ultraviolet conformal field theories.
\end{abstract}



\section{Introduction}

Treating quantum field theories in 1+1 dimensions as a test laboratory for
realistic theories in higher dimensions, this paper is concerned with the
general question of how to enlarge a given finite particle spectrum of a
theory to an infinite one.

In general the bootstrap \cite{boot}, which is the construction principle
for the scattering matrix, is assumed to close after a finite number of
steps, which means it involves a finite number of particles. However, from a
physical as well as from a mathematical point of view, it appears to be
natural to extend the construction in such a way that it would involve an
infinite number of particles. The physical motivation for this are string
theories, which admit an infinite particle spectrum. Mathematically the
infinite bootstrap would be an analogy to infinite dimensional groups, in
the sense that two entries of the S-matrix are combined into a third, which
is again a member of the same infinite set. It appears to us that it is
impossible to construct an infinite bootstrap system involving asymptotic
states and find the mathematical analogue to infinite groups in this sense
(see also footnote 2 and the paragraph after figure 1). However, it is
possible to introduce an infinite number of unstable particles into the
spectrum. Scattering matrices which would allow such type of interpretation
have occurred in the literature \cite{Karo,Z4,MP}, although only in the
latter paper a reference to unstable particles has been made. In \cite{Z4,MP}
these matrices were found to be expressible in terms of elliptic functions,
a feature very common in the context of lattice models, e.g.\cite{latt}. 
The main
purpose of this paper is to suggest a general construction principle for
such type of S-matrices starting from some known theory with a finite
particle spectrum of a special, albeit quite generic, form. As particular
examples we provide elliptic generalizations of scattering matrices related
to a pair of Lie algebras \cite{FK}, which contain the affine Toda
S-matrices \cite{TodaS} and homogeneous sine-Gordon S-matrices \cite{HSGS}
for particular choices of the algebras.

Our paper is organized as follows: In section II we provide a general \
principle for the construction of scattering matrices which involve an
infinite number of unstable particles and present some explicit examples. In
section III we construct renormalization group (RG) scaling functions by
means of the c-theorem and the thermodynamic Bethe ansatz (TBA) for the
generalization of the sinh-Gordon model. Our conclusions and an outlook
towards open problems are stated in section IV.

\section{Construction Principle}

Let us consider the huge class of two-particle S-matrices, which describe
the scattering between particles of type $a$ and $b$ as a function of the
rapidity difference $\theta $, of the general form\footnote{%
Exceptions to this factorization are for instance the scattering matrices of
affine Toda field theories related to non-simply laced Lie algebras, which
was first noted in \cite{Gust}.} 
\begin{equation}
S_{ab}(\theta )=S_{ab}^{\min }(\theta )S_{ab}^{\text{CDD}}(\theta )\,.
\label{fact}
\end{equation}
Here $S_{ab}^{\min }(\theta )$ denotes the so-called minimal S-matrix which
satisfies the consistency relations \cite{boot}, namely unitarity, crossing
and the fusing bootstrap equations and possibly possess poles on the
imaginary axis in the sheet $0\leq 
\mathop{\rm Im}%
\theta \leq \pi $, which is physical for asymptotic states. The CDD-factor 
\cite{CDD}, referred to as $S_{ab}^{\text{CDD}}(\theta )$, also satisfies
these equations, but has its poles in the sheet $-\pi \leq 
\mathop{\rm Im}%
\theta \leq 0$, which is the ``physical one'' for resonance states. $S_{ab}^{%
\text{CDD}}(\theta )$ might depend on additional constants like the
effective coupling constant or a resonance parameter. A simple prescription
to introduce now an infinite number of resonance poles is to replace the
CDD-factor in (\ref{fact}) by 
\begin{equation}
\hat{S}_{ab}^{\text{CDD}}(\theta ,N)=\prod\limits_{n=-N}^{N}S_{ab}^{\text{CDD%
}}(\theta +n\omega )\,,  \label{pr}
\end{equation}
where $\omega $ is taken to be real. By construction the new S-matrix, $\hat{%
S}_{ab}(\theta ,N)=S_{ab}^{\min }(\theta )\hat{S}_{ab}^{\text{CDD}}(\theta
,N)$ satisfies the bootstrap consistency equations and possible poles in the
sheet $-\pi \leq 
\mathop{\rm Im}%
\theta \leq 0$ have now been duplicated $2N$ times within this sheet, such
that they admit an interpretation as unstable particles. Therefore, when $%
N\rightarrow \infty $ we have an infinite number of resonance poles. Since,
as a consequence of crossing and unitarity, the S-matrix is known to be a $%
2\pi i$-periodic function, a property shared individually by $S_{ab}^{\text{%
CDD}}(\theta ,N)$, \ we expect to recover a double periodic function in the
limit $N\rightarrow \infty $ 
\begin{equation}
\lim_{N\rightarrow \infty }\hat{S}_{ab}^{\text{CDD}}(\theta
,N)=\lim_{N\rightarrow \infty }\hat{S}_{ab}^{\text{CDD}}(\theta +\mu 2\pi
i+\nu \omega ,N)  \label{prin}
\end{equation}
for $\mu ,\nu \in {\Bbb Z}$. At this stage it is not clear whether the
prescription (\ref{pr}) is meaningful at all, in the sense that it leads to
meaningful quantum field theories, and in particular one has to be concerned
about the convergence of the infinite product in (\ref{prin}). Since $%
\lim_{N\rightarrow \infty }\hat{S}_{ab}^{\text{CDD}}(\theta ,N)$ \ is a
double periodic function we expect that it is somehow related to elliptic
functions (see e.g. \cite{Elliptic} for their properties). Let us therefore
now look concretely at the building blocks which can be used to make up the
entire scattering matrix in the non-elliptic case, when backscattering is
absent. In that case the S-matrices are diagonal and known \cite{Mitra} to
be of the general form 
\begin{equation}
S_{ab}(\theta )=\prod_{x\in {\cal A}}\{x\}_{\theta }^{\sigma }\,=\prod_{x\in 
{\cal A}}\frac{\tanh (\theta -i\pi x+\sigma )/2}{\tanh (\theta +i\pi
x+\sigma )/2},  \label{gen}
\end{equation}
with $x\in {\Bbb Q}$ and $\sigma \in {\Bbb R}$. A specific theory is then
characterized by the finite set ${\cal A}$\footnote{%
The fact that $x\in {\Bbb Q}$ together with the bootstrap leads to a finite
set ${\cal A}$ and therefore a finite number of asymptotic states. Taking
instead $x\in {\Bbb R}$ could possibly lead to an infinite number, but a
consistent closure of the bootstrap is not known up to now.}. This means, if
we demonstrate that the prescription (\ref{prin}) is meaningful for each
individual building block $\{x\}_{\theta }^{\sigma }$ as defined in (\ref
{gen}), in particular we need to demonstrate the convergence of the infinite
product, we have established that it is sensible for the entire scattering
matrix. For this purpose we note the identity 
\begin{equation}
\{x\}_{\theta ,\ell }^{\sigma }:=\prod_{n=-\infty }^{\infty }\{x\}_{\theta
+n\omega }^{\sigma }=\frac{%
\mathop{\rm sc}%
\theta _{-}%
\mathop{\rm dn}%
\theta _{+}}{%
\mathop{\rm sc}%
\theta _{+}%
\mathop{\rm dn}%
\theta _{-}}.  \label{bl1}
\end{equation}
Here we abbreviated $\theta _{\pm }=(\theta \pm i\pi x+\sigma )iK_{\ell
}/\pi $ and used the Jacobian elliptic functions in the standard notation $%
\mathop{\rm pq}%
(z)$ with p,q $\in \{$s,c,d,n$\}$ (see e.g. \cite{Elliptic}). The quarter
periods $K_{\ell }$ depending on the parameter $\ell \in \lbrack 0,1]$ are
defined in the usual way through the complete elliptic integral 
\begin{equation}
K_{\ell }=\int_{0}^{\pi /2}(1-\ell \sin ^{2}\theta )^{-1/2}d\theta \,.
\end{equation}
The period of $\{x\}_{\theta ,\ell }^{\sigma }$ is chosen to be $\omega =\pi
K_{(1-\ell )}/K_{\ell }$. The last identity in (\ref{bl1}) is easily derived
from the infinite product representations of the elliptic functions which
can be found in various places as for instance in \cite{Elliptic} 
\begin{eqnarray}
\mathop{\rm sc}%
x &=&k\tan \frac{\pi x}{2K_{\ell }}\prod_{n=1}^{\infty }\frac{1-2q^{2n}\cos
(\pi x/K_{\ell })+q^{4n}}{1+2q^{2n}\cos (\pi x/K_{\ell })+q^{4n}}, \\
\mathop{\rm dn}%
x &=&k^{-1}\prod_{n=1}^{\infty }\frac{1+2q^{2n-1}\cos (\pi x/K_{\ell
})+q^{4n-2}}{1-2q^{2n-1}\cos (\pi x/K_{\ell })+q^{4n-2}}\,\,,
\end{eqnarray}
with $k=(1-\ell )^{-1/4}$ and $q=\exp (-\omega )$. Recalling the well known
limits $\lim_{\ell \rightarrow 0}K_{\ell }=\pi /2$ and $\lim_{\ell
\rightarrow 0}K_{(1-\ell )}=\infty $ we obtain 
\begin{equation}
\lim_{\ell \rightarrow 0}\{x\}_{\theta ,\ell }^{\sigma }=\{x\}_{\theta
}^{\sigma }\,.  \label{aa}
\end{equation}
This means in the limit $\ell \rightarrow 0$ the elliptic S-matrix $\hat{S}%
_{ab}(\theta )$ collapses to the hyperbolic one, that is $S_{ab}(\theta )$.
Notice that due to the general identity $%
\mathop{\rm pr}%
(x)/%
\mathop{\rm qr}%
(x)=%
\mathop{\rm pq}%
(x)$, we could also write (\ref{bl1}) in terms of various other combinations
of elliptic functions. For instance replacing $%
\mathop{\rm sc}%
$ by $%
\mathop{\rm sn}%
/%
\mathop{\rm cn}%
$ is probably most intuitive, since it allows an alternative prescription to
(\ref{prin}) for the construction of elliptic scattering matrices: Replace $%
\sinh \rightarrow 
\mathop{\rm sn}%
$, $\cosh \rightarrow 
\mathop{\rm cn}%
$ and correct the consistency equations by a factor $%
\mathop{\rm dn}%
$, which reduces always to $1$ in the hyperbolic limit, in such a way that
no resonance poles are left inside the physical sheet. Defining now the
function 
\begin{equation}
\theta _{\mu ,\nu }(x,\sigma ,\omega ):=2\pi i(\nu +x/2)+2\omega \mu -\sigma
\end{equation}
the singularities of $\{x\}_{\theta ,\ell }^{\sigma }$ are easily identified
as 
\begin{eqnarray}
\text{zeros} &:&\text{\quad }\theta _{\mu ,\nu }(x,\sigma ,\omega ),\quad
\,\,\,\,\,\theta _{\mu ,\nu }(1-x,\sigma ,\omega ), \\
\text{poles} &:&\text{\quad }\theta _{\mu ,\nu }(-x,\sigma ,\omega ),\quad
\theta _{\mu ,\nu }(x-1,\sigma ,\omega )\,\,.
\end{eqnarray}
Note that when taking $0\leq x\leq 1$ the poles are situated in the
non-physical sheet.

This brings us to the question of how to interpret these poles and how can
we characterize the physical properties of the related particles?
Considering the S-matrix $S_{ab}(\theta )$, which describes the scattering
of two particles of type $a\,$and $b$ with masses $m_{a}$ and $m_{b}$, we
assume that there is a resonance pole situated at $\theta _{R}=\sigma -i\bar{%
\sigma}$. According to the Breit-Wigner formula \cite{BW} (see also e.g. 
\cite{ELOP}) the mass $M_{\tilde{c}}$ and the decay width $\Gamma _{\tilde{c}%
}$ of an unstable particle of type $\tilde{c}$ can be conveniently expressed
as 
\begin{equation}
2M_{\tilde{c}}^{2}=\sqrt{\gamma ^{2}+\tilde{\gamma}^{2}}+\gamma ,\quad
\Gamma _{\tilde{c}}^{2}/2=\sqrt{\gamma ^{2}+\tilde{\gamma}^{2}}-\gamma ,
\label{BrW}
\end{equation}
where 
\begin{eqnarray}
\gamma &=&m_{a}^{2}{}+m_{b}^{2}{}+2m_{a}m_{b}\cosh \sigma \cos \bar{\sigma}%
,\,\,  \label{BW2} \\
\tilde{\gamma} &=&2m_{a}m_{b}\sinh |\sigma |\sin \bar{\sigma}\,.
\end{eqnarray}
We keep here in mind that this description, although frequently used e.g. 
\cite{HSGS,MP,CF2}, is not entirely rigorous and requires additional
investigation. This caution is based on various facts. First, the relations (%
\ref{BrW}) are simply derived by carrying over a prescription from usual
quantum mechanics to quantum field theory, i.e. complexifying the mass of a
stable particle. Second, solving the Breit-Wigner formula for the quantities 
$M_{\tilde{c}}$, $\Gamma _{\tilde{c}}$ and treating them literally as mass
and decay width is somewhat problematic since this is in conflict with
Heisenberg's uncertainty principle, because apparently we know
simultaneously the energy and the time. Third, the Breit-Wigner relations
presume an exponential decay in momentum space, which is in fact
incompatible with the general principles of quantum field theory and
therefore might possibly be a problem in this context \cite{Bert}.
Nonetheless, we employ these quantities and try to find evidence to support
that they are indeed meaningful. When taking $M_{\tilde{c}}$ to be the mass
of the unstable particle there should be a threshold for energetic reasons
of the type 
\begin{equation}
M_{\tilde{c}}\geq m_{a}+m_{b}\,,  \label{th1}
\end{equation}
with the consequence that the decay width is bounded by 
\begin{equation}
\Gamma _{\tilde{c}}^{2}\geq 8m_{a}m_{b}(1-\cosh \sigma \cos \bar{\sigma}%
)\,\,.  \label{th2}
\end{equation}
So far evidence for these thresholds has not been found in the literature.
One reason for this is that the unstable particles enter the bootstrap
principle in a more passive way than the stable particles, whose properties
are directly used in the construction procedure. Hence one expects that
signs for these thresholds will emerge in a more indirect way.

A summary of our statements about the pole structure of $\hat{S}(\theta )$
is depicted in figure 1.

\begin{center}
\includegraphics[width=8.2cm,height=6.09cm]{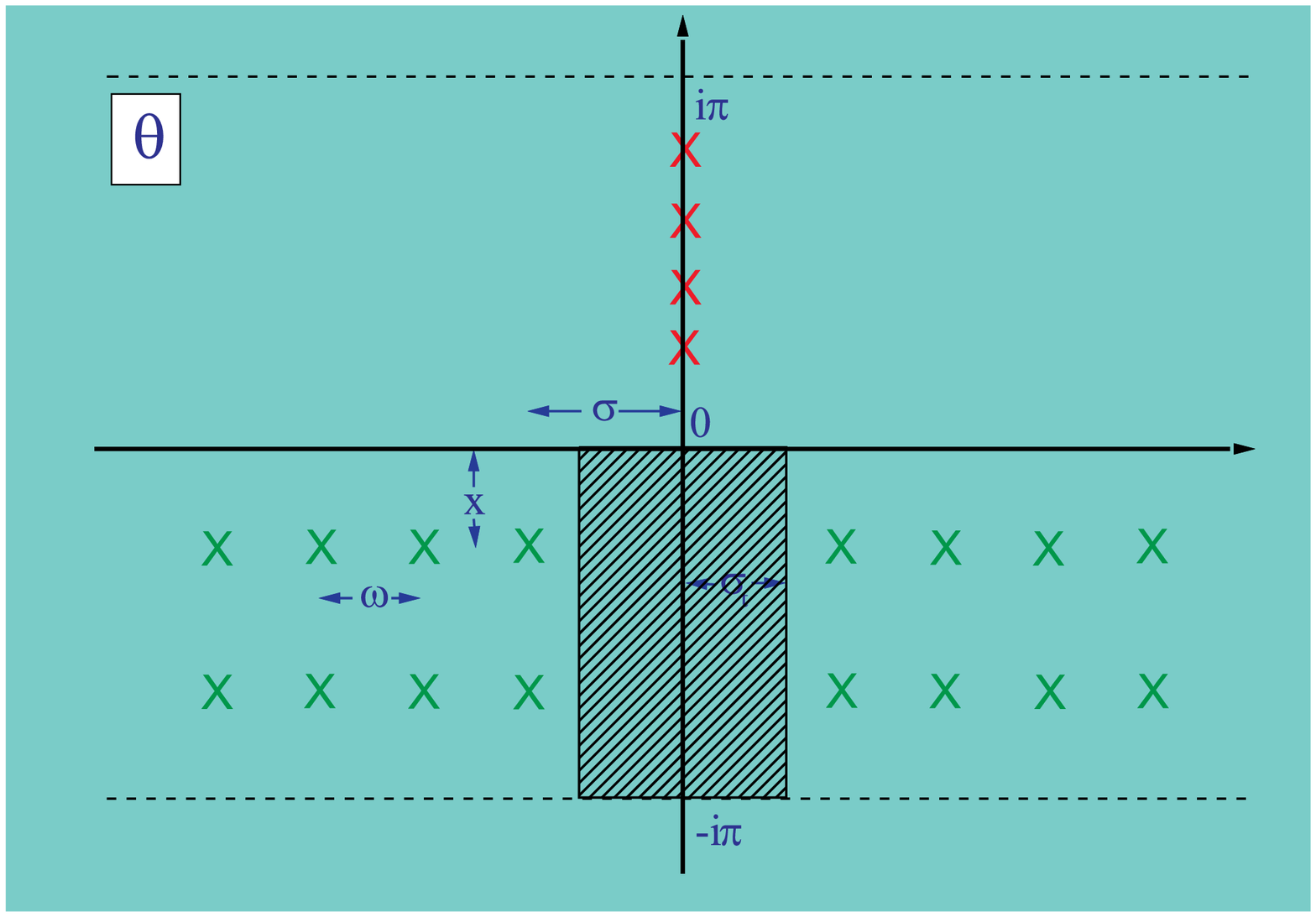}
\end{center}

\vspace*{0.2cm}

\noindent {\small Figure 1: The poles of the blocks }$\{x\}_{\theta ,\ell
}^{\sigma }$ {\small are the crosses in the sheet }$-\pi \leq 
\mathop{\rm Im}%
\theta \leq 0$. {\small The crosses on the positive part of the imaginary
axis are associated, as usual, with stable particles. For equal masses of
the stable particles the threshold (\ref{th1}) is }$\sigma _{t}=%
\mathop{\rm arccosh}%
[(3-\cos \pi x)/(1+\cos \pi x)].\medskip ${\small \ }

Since the poles inside the sheet $0\leq 
\mathop{\rm Im}%
\theta \leq \pi $ are associated to $S_{ab}^{\min }(\theta )$, it is also
obvious from figure 1 why the prescription (\ref{pr}) may not be employed
for this part of the S-matrix, since it would lead to a pole structure which
is, according to (\ref{BrW}), non physical for $M_{\tilde{c}}$ and $\Gamma _{%
\tilde{c}}$.

\subsection{Examples}

It is clear by construction that our prescription includes all affine Toda
field theories related to simply laced Lie algebras, since they all
factorize as (\ref{fact}) and may be represented in the form (\ref{gen}),
see e.g. \cite{TodaS}. Taking the resonance parameter $\sigma $ to be zero
and the set ${\cal A}=\{t\}$ for $0\leq t\leq 1$, we recover as a special
case the elliptic version of the sinh-Gordon model proposed in \cite{MP}.
Reintroducing $\sigma $, its scattering matrix reads 
\begin{equation}
\hat{S}(\theta )=\prod\limits_{n=-\infty }^{\infty }\frac{\tanh (\theta
-i\pi x+n\omega +\sigma )/2}{\tanh (\theta +i\pi x+n\omega +\sigma )/2}\,.
\label{ESG}
\end{equation}
According to (\ref{BrW}) the masses and decay width of the unstable
particles are 
\begin{eqnarray}
M_{\mu ,\nu }^{\sigma ,n\omega } &=&m\sqrt{2}\cosh \frac{\theta _{\mu ,\nu
}(y,\sigma ,n\omega )}{2},\,\,\,\,\,\,y=-x,x-1 \\
\Gamma _{\mu ,\nu }^{\sigma ,n\omega } &=&m2\sqrt{2}\sinh \frac{\theta _{\mu
,\nu }(y,\sigma ,n\omega )}{2},\,\,\,\,y=-x,x-1
\end{eqnarray}
where $m$ denotes the mass of the stable particle. The thresholds (\ref{th1}%
) and (\ref{th2}) translate in this case into 
\begin{equation}
\cosh (n\omega +\sigma )\geq \frac{3\mp \cos \pi x}{1\pm \cos \pi x},\Gamma
\geq 4m\frac{\sin ^{2}\frac{\pi (2x+1\pm 1)}{4}}{\cos \frac{\pi (2x+1\pm 1)}{%
4}}.  \label{tresh}
\end{equation}
As a further example we consider the elliptic generalization of the $%
A_{1}|A_{N-1}$-theory ($\equiv SU(N)_{2}$-homogeneous sine-Gordon model).
The two-particle S-matrix describing the scattering of two stable particles
of type $a$ and $b$, with $1\leq a,b\leq N-1$, related to the non-elliptic
version of this model was proposed in \cite{HSGS}. In our notation it may be
written as 
\begin{equation}
S_{ab}(\theta ,\sigma _{ab})=(-1)^{\delta _{ab}}\left[ c_{a}\sqrt{%
\{1/2\}_{\theta }^{\sigma _{ab}}}\,\right] ^{I_{ab}}\,.  \label{ZamS}
\end{equation}
Here $I$ denotes the incidence matrix of the $SU(N)$-Dynkin diagram, the
resonance parameters have the property $\sigma _{ab}=-\sigma _{ba}$ and $%
c_{a}=\pm 1$ depending on whether $a$ is even or odd. According to our
prescription outlined in the previous paragraph, the elliptic generalization
of (\ref{ZamS}) is 
\begin{equation}
\hat{S}_{ab}(\theta ,\sigma _{ab},\ell )=(-1)^{\delta _{ab}}\left[ c_{a}%
\sqrt{\{1/2\}_{\theta ,\ell }^{\sigma _{ab}}}\,\right] ^{I_{ab}}\,.
\end{equation}
Note that despite the appearance of the square root, $S$ as well as $\hat{S}$
are still meromorphic functions in $\theta $.

\section{RG-scaling functions}

Having established that our prescription leads to sensible solutions of the
bootstrap consistency equations, we would also like to know what kind of
quantum field theories these scattering matrices correspond to. Up to now
all known solutions to the on-shell consistency equations have led to
sensible QFT's, albeit a rigorous proof which would establish that indeed 
{\it all }solutions are well-defined local QFT's is still an outstanding
issue. Some crucial characteristics of the theory are contained in the
renormalization group scaling functions, which we now want to determine. In
particular, we want to identify in the extreme ultraviolet limit the
Virasoro central charges of the corresponding conformal field theories.

\subsection{The c-theorem}

We carry out this task by evaluating the c-theorem \cite{ZamC} in the
version presented in \cite{CF2} 
\begin{eqnarray}
&&c(r)=3\sum_{n=1}^{\infty }\sum_{\mu _{1}\ldots \mu
_{n}}\int\limits_{-\infty }^{\infty }\frac{d\theta _{1}\ldots d\theta _{n}}{%
n!(2\pi )^{n}}e^{-r\,E}  \label{cr0} \\
&&\times \left| F_{n}^{\Theta |\mu _{1}\ldots \mu _{n}}(\theta _{1},\ldots
,\theta _{n})\right| ^{2}\frac{(6+6rE+3r^{2}E^{2}+r^{3}E^{3})}{2E^{4}}\,. 
\nonumber
\end{eqnarray}
The sum of the on-shell energies is here denoted by $E=\sum%
\nolimits_{i=1}^{n}m_{\mu _{i}}\cosh \theta _{i}$, with $m_{\mu _{i}}$ being
the masses of the theory and the correlation function for the trace of the
energy momentum tensor $\Theta $ has been expanded in terms of $n$-particle
form factors $F_{n}^{\Theta |\mu _{1}\ldots \mu _{n}}(\theta _{1},\ldots
,\theta _{n})$ (see \cite{Kar} for general properties and \cite{FMS} for
explicit sinh-Gordon formulae). We normalized $\Theta $ and $m_{\mu _{i}}$
by an overall mass scale, such that $E$ as well as the renormalization group
parameter $r$ become dimensionless. In particular $\lim_{r\rightarrow 0}c(r)$
is the ultraviolet Virasoro central charge.

Let us now start with the evaluation of $c(r)$ as defined in (\ref{cr0}) for
the elliptic version of the sinh-Gordon model. As the input for this we need
to know the $n$-particle form factors. Since so far it is not known how to
compute the sum in $n$ analytically, we have to resort to a numerical
treatment and it is clear that we have to terminate the series at a certain
value of $n$. Fortunately, it was observed explicitly in \cite{FMS}, that in
fact the expression for $n=2$ is already very close to the exact answer for
the sinh-Gordon model. We assume here that the convergence behaviour is
still true when we generalize the scattering matrix to (\ref{ESG}). Note
that in general one has to be careful with this approximation, since the
higher particle contributions are crucial in some models in order to obtain
a good approximation to $c(r)$ \cite{CF1,CF2,CF3}. In the two-particle
approximation, indicated by the superscript, one can perform one of the
integrations analytically and (\ref{cr0}) acquires the simple form 
\begin{equation}
\lim_{r\rightarrow 0}c^{(2)}(r)=\frac{3}{2}\int\limits_{0}^{\infty }d\theta 
\frac{|F_{2}^{\Theta }(2\theta )|^{2}}{\cosh ^{4}\theta }\,.
\label{thefamouskink}
\end{equation}
It is here crucial to note that besides the formulation of $\hat{S}$ in
terms of elliptic functions for $N\rightarrow \infty $, it can also be
expressed equivalently in terms of the usual sinh-Gordon S-matrix (\ref{bl1}%
). When trying to solve now the form factor consistency equations \cite{Kar}%
, we can exploit this observation. Since for the model at hand there is
neither a kinematic nor a bound state pole in $F_{2}^{\Theta }(\theta )$,
the only equations to be solved are Watson's equations. The two particle
form factor is then easily obtained to be 
\begin{equation}
\hat{F}_{2}^{\Theta }(\theta ,N)=2\pi \prod\limits_{n=-N}^{N}\text{ }\frac{%
F_{\text{min}}(\theta +n\omega )}{F_{\text{min}}(i\pi +n\omega )},
\label{coj}
\end{equation}
where $F_{\text{min}}(\theta )$ is the minimal form factor of the
sinh-Gordon model obtained in \cite{FMS} 
\begin{eqnarray}
F_{\text{min}}(\theta ) &=&\exp \left[ 4\int\limits_{0}^{\infty }\frac{dt}{t}%
\,\left( \cos \left( \frac{t\theta }{\pi }\right) \coth t+i\sin \left( \frac{%
t\theta }{\pi }\right) \right) \right.  \nonumber \\
&&\times \left. \frac{\sinh (\frac{t(x-1)}{2})\sinh (\frac{tx}{2})\sinh (%
\frac{t}{2})}{\sinh (t)}\right] \,.  \label{fe}
\end{eqnarray}
Using the infinite product representation for $F_{\text{min}}(\theta )$ \cite
{FMS} the solution (\ref{coj}) for $N\rightarrow \infty $ coincides with the
equation (5.3) in \cite{MP}. Proceeding now to the evaluation of (\ref{cr0}%
), we require $|\hat{F}_{2}^{\Theta }(\theta ,N)|^{2}$, whose
characteristics are captured in figure 2.

\begin{center}
\includegraphics[width=8.2cm,height=6.09cm]{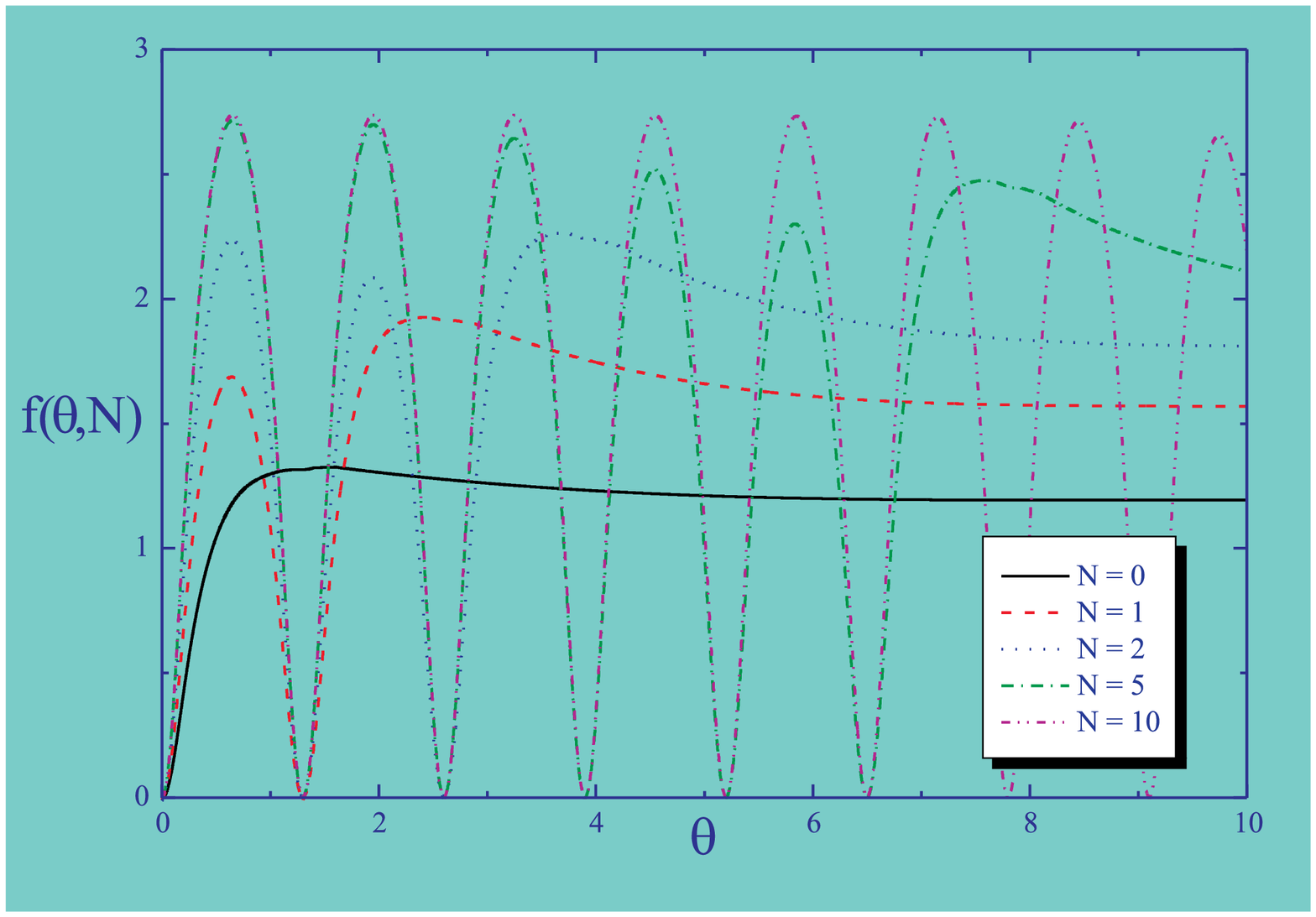}
\end{center}

\vspace*{0.2cm}

\noindent {\small Figure 2: Absolute value squared of the two particle form
factors }f$(\theta ,N)=|\hat{F}_{2}^{\Theta }(\theta ,N)/2\pi |^{2}$ {\small %
as functions of the rapidity for different values of }$N${\small \ for $%
\omega =1.3$ and $x=0.1$.\medskip }

We observe that for a certain value of $N$ the function starts to converge,
which is of course important from a technical point of view when we want to
compute the limiting case $N\rightarrow \infty $. The other observation we
make in figure 2, see also figure 5 in \cite{FMS}, is that f$(\theta ,N=0)$
always has a distinct maximum, which we refer to as $\theta _{m}$. From (\ref
{fe}) follows that it is determined by the solution of 
\begin{eqnarray}
&&\frac{4\theta _{m}}{\pi }\cosh \theta _{m}\sin \pi x+\cosh 2x\pi \coth 
\frac{\theta _{m}}{2}  \nonumber \\
&=&\frac{\cosh \frac{3\theta _{m}}{2}}{\sinh \frac{\theta _{m}}{2}}%
+2(2x-1)\cos \pi x\sinh \theta _{m}\,.  \label{for HB}
\end{eqnarray}
Solving this equation for various values of $x$, we find that $\theta _{m}$
is always slightly greater than the smallest threshold bound obtained from (%
\ref{tresh}). For instance for $x=0.1$ we obtain $\theta _{m}\simeq 1.439$
and $n\omega +\sigma >0.315$ and for $x=0.5$ we have $\theta _{m}\simeq
2.040 $ and $n\omega +\sigma >1.763$. We interpret this as an indication
that the form factors ``know'' about the thresholds (\ref{tresh}). We
support this now by considering \ $\lim_{N\rightarrow \infty }$ f$(\theta ,N)
$ for various values of $\omega $.

\begin{center}
\includegraphics[width=8.2cm,height=6.09cm]{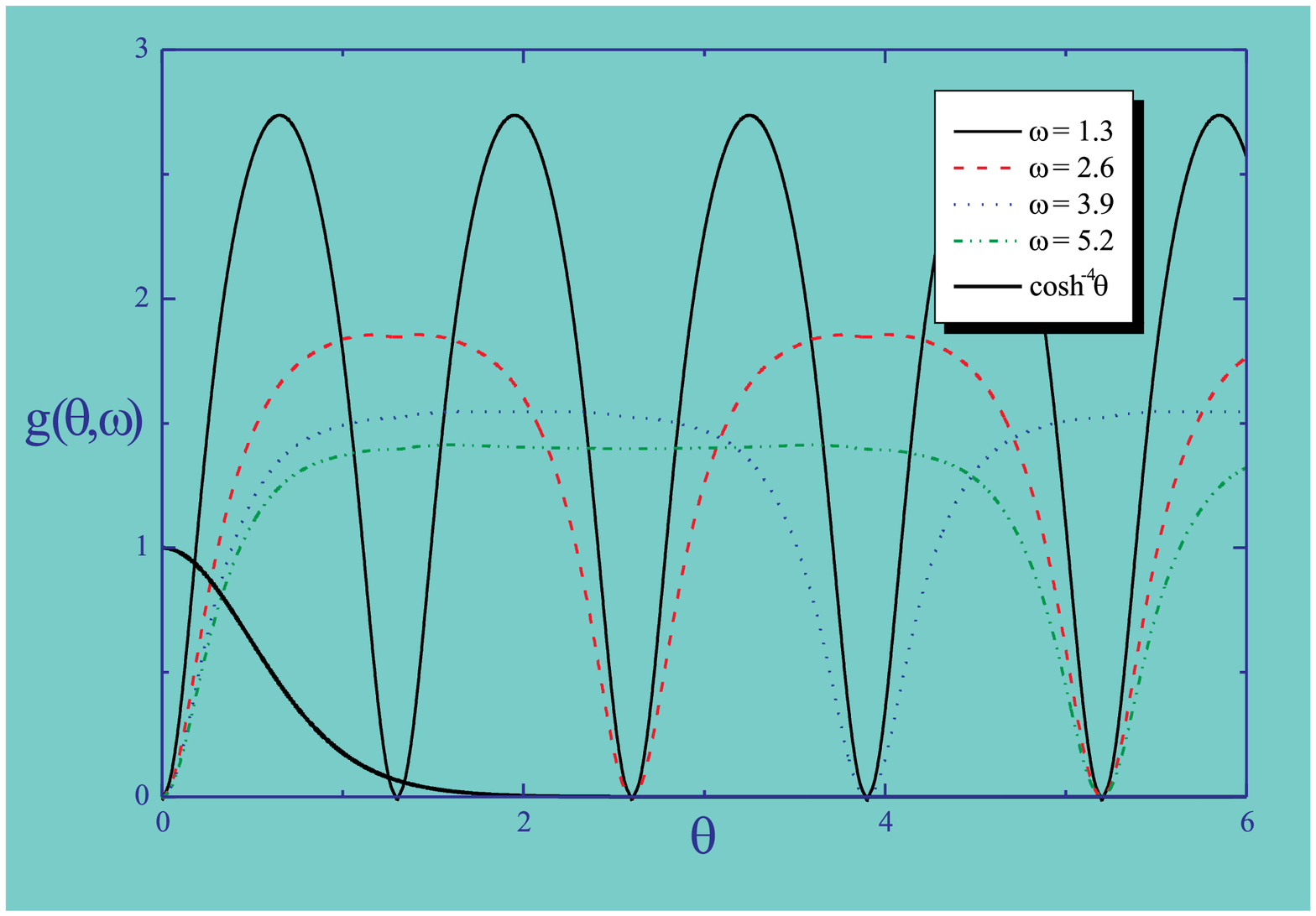}
\end{center}

\vspace*{0.2cm}

\noindent {\small Figure 3: Absolute value squared of minimal form factors }g%
$(\theta ,\omega )=\lim_{N\rightarrow \infty }|\hat{F}_{\text{min}}(\theta
,N)|^{2}$ {\small as functions of the rapidity for different values of $%
\omega $ and $x=0.1$}$.\medskip $

\vspace*{0.2cm}

\begin{center}
\includegraphics[width=8.2cm,height=6.09cm]{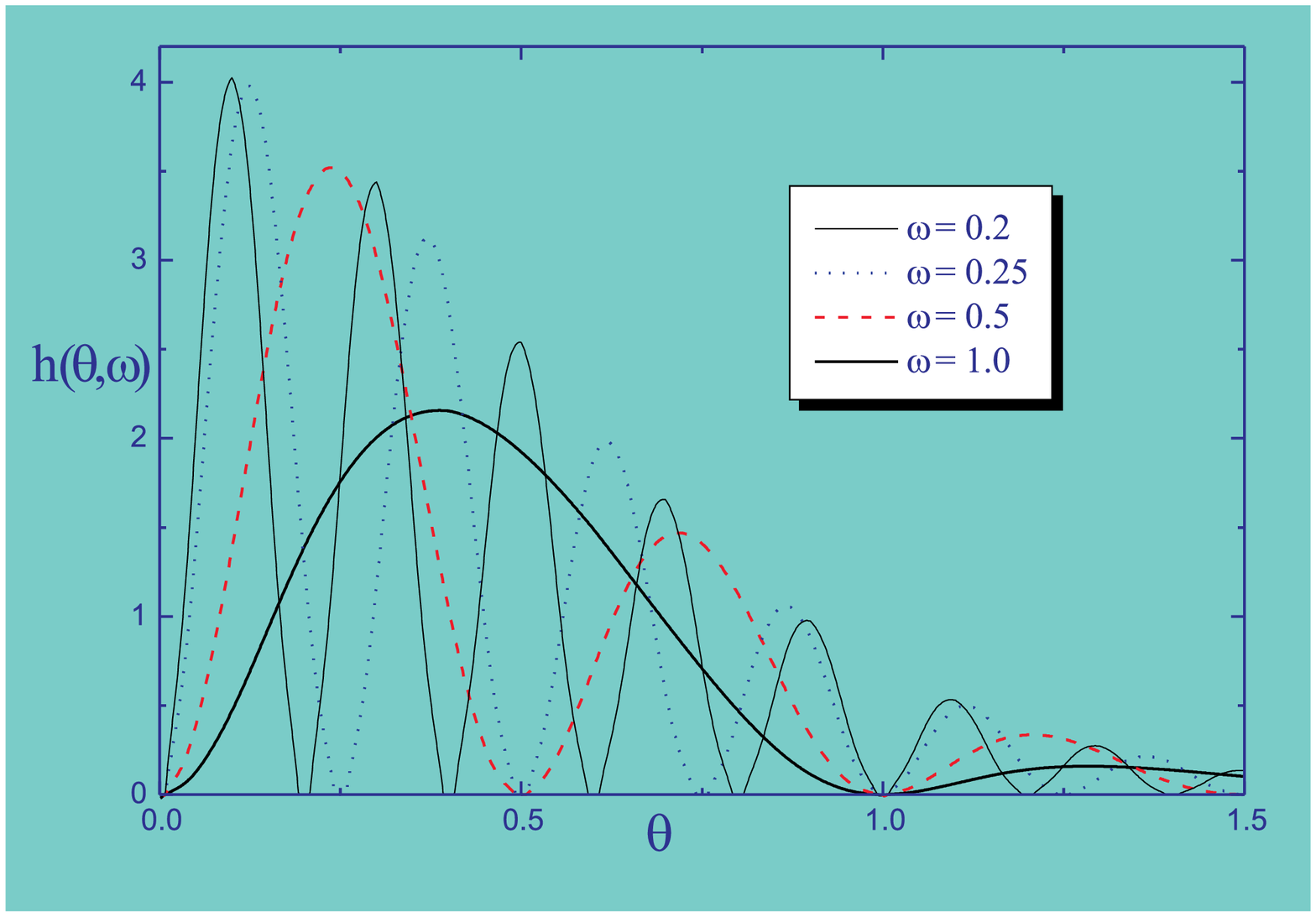}
\end{center}

\vspace*{0.2cm}

\noindent {\small Figure 4: Integrand of equation (\ref{thefamouskink}),
that is }h$(\theta ,\omega )$ $=$ $\lim_{N\rightarrow \infty }$ $|\hat{F}_{%
\text{min}}(\theta ,N)|^{2}/\cosh ^{4}\theta $ {\small as a function of the
rapidity for different values of $\omega $ and $x=0.1$.\medskip }

We observe that in the region in which the factor $1/\cosh ^{4}(\theta )$,
emerging in (\ref{thefamouskink}), is still non vanishing the integrals $%
\lim_{N\rightarrow \infty }$ $\int d\theta |\hat{F}_{\text{min}}(\theta
,N)|^{2}$ are decreasing functions of $\omega $. This behaviour is changed
once we take $\omega <\theta _{m}$ as we can explicitly extract from figure
4.

Naturally this features are also reflected in the scaling functions.
Presuming that for each value of $N$ we have a consistent theory, we would
like to know which ultraviolet central charges these models possess and in
addition we want to identify a value of $N$ for which the related model
constitutes reasonably good approximation for the elliptic models. That such
an identification is possible is exhibited in figure 5. In addition we
observe, that for fixed $\omega $ and $x$ the scaling function is a
monotonically increasing when $N$ is varied.

\begin{center}
\includegraphics[width=8.2cm,height=6.09cm]{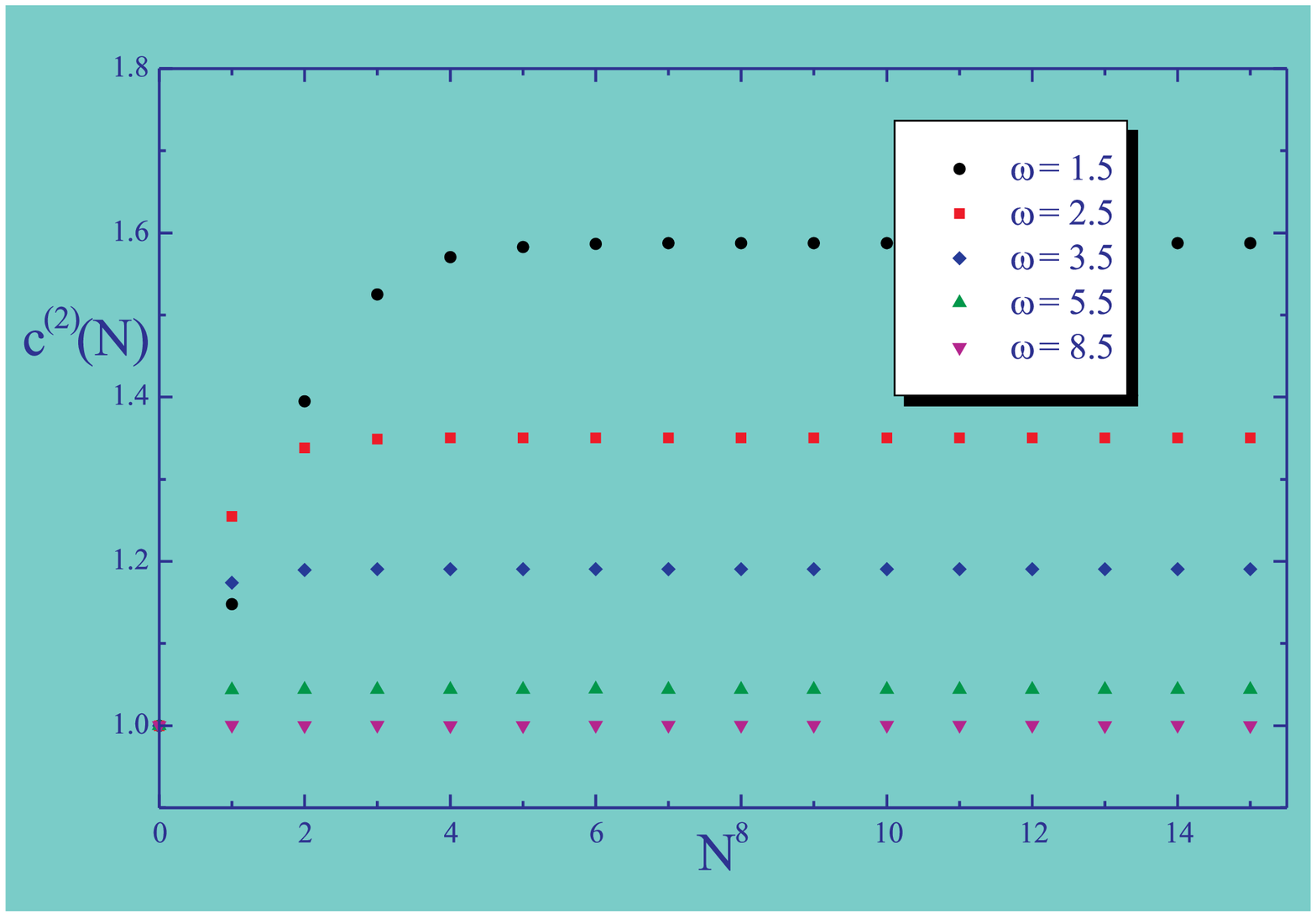}
\end{center}

\noindent {\small Figure 5: Ultraviolet Virasoro central charge }$c^{(2)}$%
{\small \ as a function of }$N${\small .\medskip }

Focussing now on the elliptic case, that is we select a large enough $N$
such that this case is well approximated, we compute the scaling function in
dependence of $\omega $ for various values of $r$. Our results are depicted
in figure 6.

\begin{center}
\includegraphics[width=8.2cm,height=6.09cm]{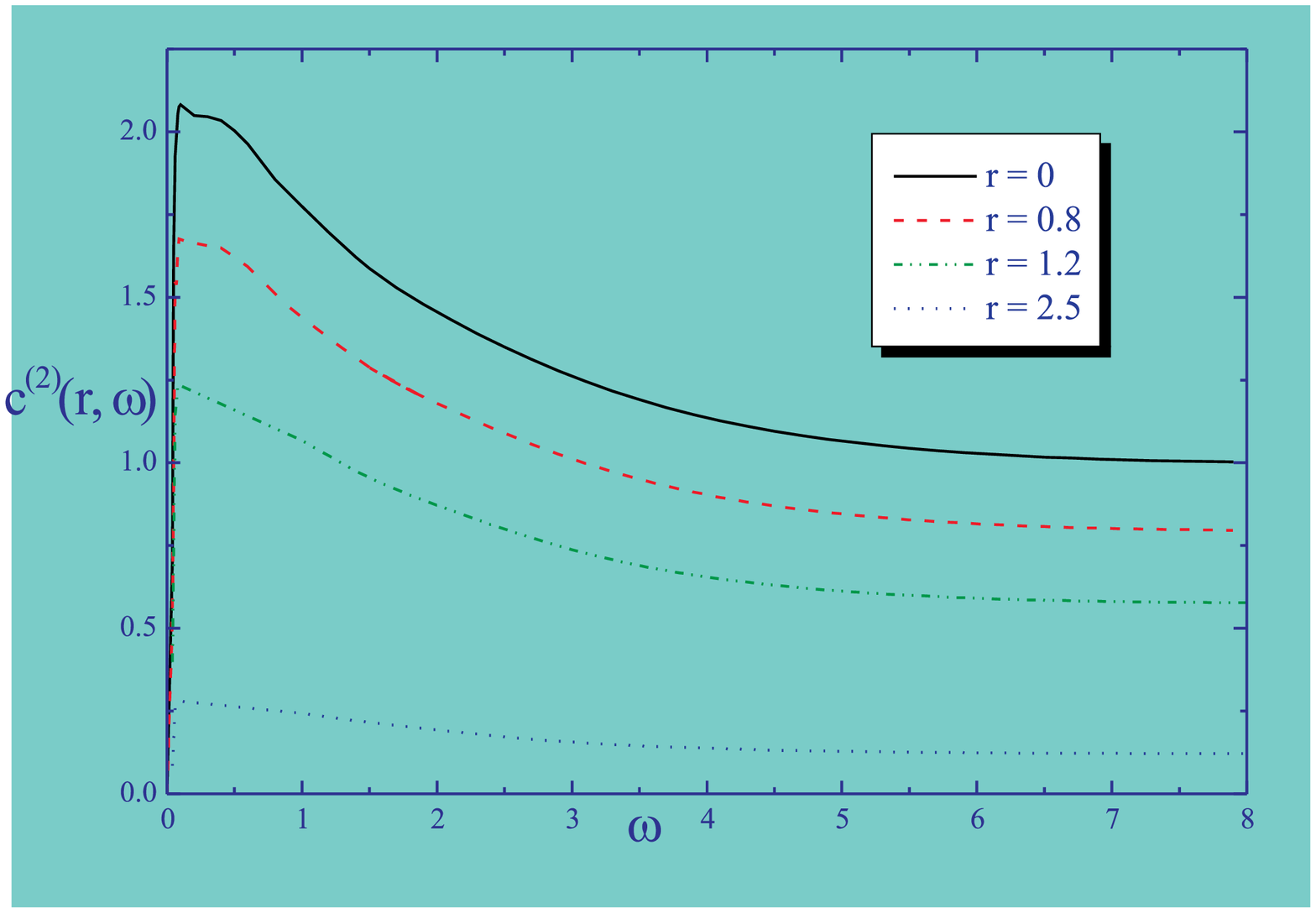}
\end{center}

\vspace*{0.2cm}

\noindent {\small Figure 6: Ultraviolet Virasoro central charge }$c^{(2)}$%
{\small \ as a function of $\omega $. } \medskip

In the extreme limits we obtain $\lim_{\omega \rightarrow 0}$ $c(r,\omega
)=0 $ and for large $\omega $ we recover the values of the sinh-Gordon
model, $\lim_{\omega \rightarrow \infty }c(r,\omega )=c_{\text{SG}}(r)$. The
latter limit follows from (\ref{aa}) and is in addition compatible with $%
\lim_{\theta \rightarrow \infty }F_{\text{min}}(\theta )=1$. The values for
the extremal points were already quoted in \cite{MP}, however, we also
observe that the function is not monotonically increasing between these
points as claimed in there. In fact, in the physical region, that is for
values of $\omega >0.315,$ the function is monotonically decreasing and does
not take on values between $0$ and $1$. Remarkably, the threshold is quite
clearly exhibited by a drastic change in the behaviour of $c$, that is the
onset of a small plateau as is visible in figure 6. We performed the same
computation for different values of $x$ and observed that this onset moves
in the direction predicted by equation (\ref{tresh}).

\subsection{The thermodynamic Bethe ansatz}

Let us now compare the results of the previous section with an alternative
method, namely the thermodynamic Bethe ansatz \cite{TBAZam}. For this we
first have to solve the TBA-equations 
\begin{equation}
r\,\hat{m}_{i}\cosh \theta +\ln (1-e^{-L_{i}(\theta
)})=\sum\limits_{j}\,\,\varphi _{ij}\ast L_{j}(\theta )  \label{tba}
\end{equation}
for the function $L_{i}(\theta )$. The information of the scattering matrix
is captured in the kernel $\,\varphi _{ij}(\theta )=-id\ln S_{ij}(\theta
)/d\theta $ of the rapidity convolution, which is denoted as usual by $f\ast
g(\theta ):=\int d\theta ^{\prime }/2\pi \,f(\theta -\theta ^{\prime
})g(\theta ^{\prime })$. The dimensionless parameter $r=m_{1}T^{-1}$ is the
inverse temperature $T$ times the overall mass scale of the lightest
particle $m_{1}$. Also all masses have been normalized in this way, i.e. $%
\hat{m}_{i}=m_{i}/m_{1}$. Having determined the $L_{i}(\theta )$-functions,
we may compute the scaling function by means of 
\begin{equation}
c^{\prime }(r)=\frac{3\,r}{\pi ^{2}}\sum_{i}\hat{m}_{i}\int\nolimits_{0}^{%
\infty }d\theta \,\cosh \theta \,L_{i}(\theta )\,\,\,.  \label{scale}
\end{equation}
Once again $\lim_{r\rightarrow 0}c^{\prime }(r)$ is the ultraviolet Virasoro
central charge. We would like to recall here that the scaling functions $%
c(r) $ and $c^{\prime }(r)$ are not identical, but contain qualitatively the
same information in the RG sense.

In order to carry out this analysis we need to know in (\ref{tba}) the
kernel $\varphi (\theta )$ as input. For the model under consideration we
can exploit the factorization property (\ref{ESG}) for a finite product and
trivially obtain 
\begin{equation}
\varphi _{N}(\theta )=\sum\limits_{n=-N}^{N}\varphi _{\text{SG}}(\theta
+n\omega +\sigma )\,,
\end{equation}
where $\varphi _{\text{SG}}(\theta )$ is the sinh-Gordon kernel, e.g. \cite
{FKS} 
\begin{equation}
\varphi _{\text{SG}}(\theta )=\frac{4\sin (\pi x)\cosh \theta }{\cosh
(2\theta )-\cos (2\pi x)}\,.
\end{equation}
Using alternatively the representation of the S-matrix (\ref{ESG}) in terms
of elliptic functions, we compute the kernel directly to 
\begin{equation}
\lim\limits_{N\rightarrow \infty }\varphi _{N}(\theta )=\frac{K_{\ell }}{\pi 
}\sum\limits_{k=-,+}\left[ \frac{%
\mathop{\rm dc}%
\theta _{k}}{%
\mathop{\rm sn}%
\theta _{k}}+\ell (1-\ell )\frac{%
\mathop{\rm sn}%
\theta _{k}}{%
\mathop{\rm dc}%
\theta _{k}}\right] \,\,.  \label{beauty}
\end{equation}
With these expression we carry out our numerical analysis, that is we solve
iteratively the equation (\ref{tba}) and evaluate (\ref{scale}) thereafter.
The results of this investigations are presented in figure 7.

Unfortunately for very small values of $\omega $ and $r$ our numerical
iteration procedure does not converge reliably. However, we will be content
at this stage with the data obtained so far, since they already support
qualitative our c-theorem analysis. They confirm that above threshold the
scaling function is monotonically decreasing as a function of $\omega $ and
also that values greater than 1 may be reached, even for finite values of $r$%
.

\begin{center}
\includegraphics[width=8.2cm,height=6.09cm]{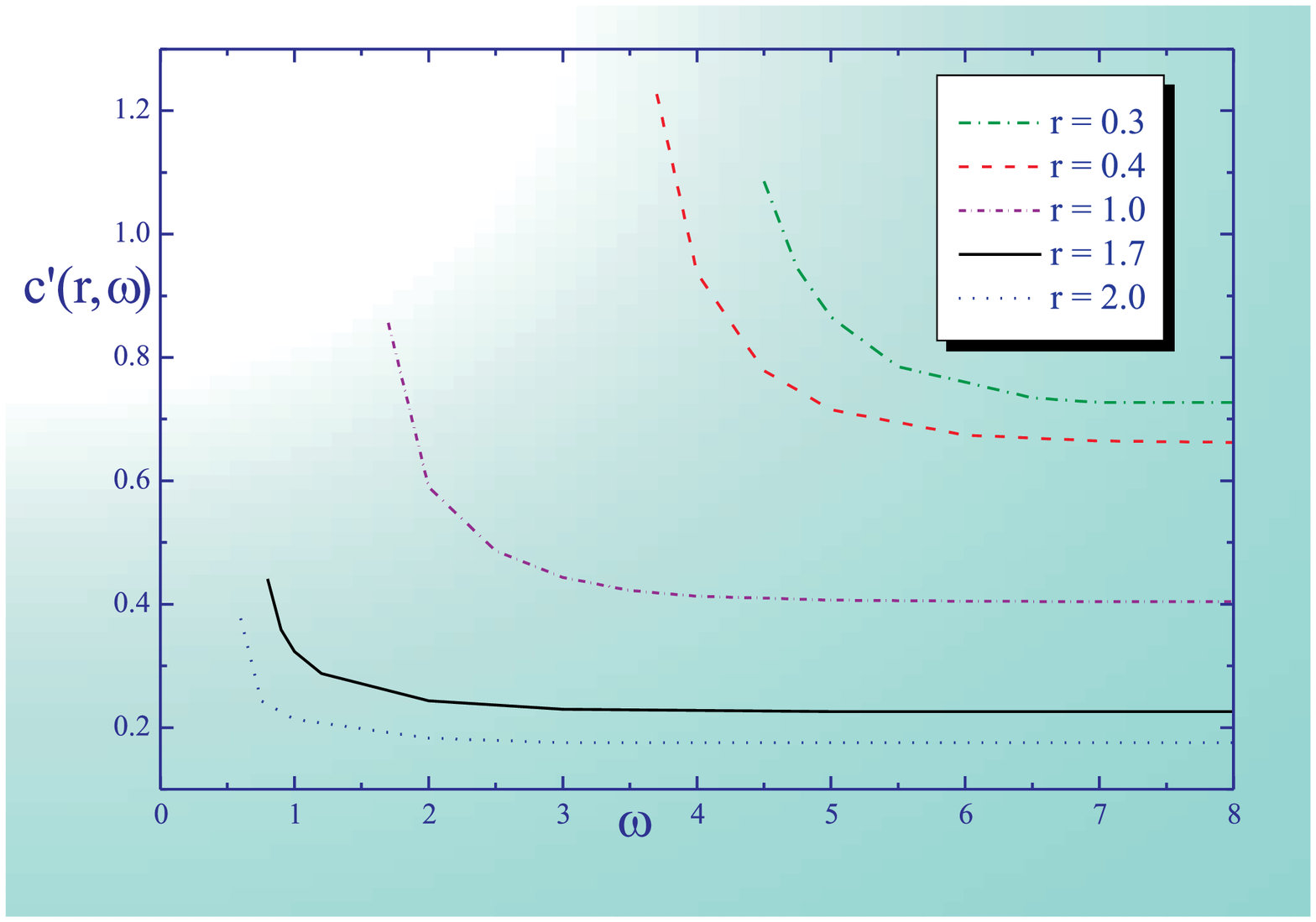}
\end{center}

\vspace*{0.2cm}

\noindent {\small Figure 7: TBA scaling function.} \medskip

\section{Conclusions}

Starting from a given scattering matrix of hyperbolic type, we have
demonstrated that it is possible to include consistently an arbitrary number
of unstable particles into the spectrum of the theory. In particular when
this number becomes infinite the S-matrix may be expressed in terms of
elliptic functions.

For the generalization of the sinh-Gordon model we computed RG scaling
functions. Within these analysis we found clear evidence for the thresholds
which constrain the masses of the unstable particles. Above threshold, the
values the ultraviolet Virasoro central charges may take are between 1 and 2
(possibly slightly greater than 2) and not between 0 and 1 as suggested in 
\cite{MP}. The theories are consistent for each finite value of $N$. For
fixed resonance parameters $\omega $ and $\sigma $ the scaling functions are
non-decreasing for increasing $N$.

Concerning the investigation of the c-theorem, it would be desirable to
refine the analysis. In particular one should include higher $n$-particle
form factors into the expansion. For the elliptic version some of them were
already presented in \cite{MP}, but in general it remains a challenge to
find closed expressions for arbitrary particle numbers. At present the TBA
analysis is the least conclusive exploration and deserves further
consideration in future. In particular the regions of $\omega $ and $r,$
which were not accessible to us, should be explored and might possibly lead
to a further more concrete indication of the thresholds also in this
context. In regard to this, it will be useful develop existence criteria for
the solution of the TBA equations analogue to the one derived in \cite{FKS}.
The one presented in there can not be taken over directly, since it makes
use of the fact that $\int d\theta |\varphi (\theta )|$ equals $2\pi $,
whereas for the model investigated here this is $2\pi N$. It would be
desirable to develop analytic approximations for the TBA solutions in the
extreme ultraviolet limit, i.e. $r=0$, similar to the ones already existing
for theories with different characteristic features. \medskip

{\bf Acknowledgments:} We are grateful to the Deutsche
Forschungsgemeinschaft (Sfb288) for financial support and to J.~Drei\ss ig
and M.~M\"{u}ller for discussions.

\end{document}